\documentclass[a4paper,12pt]{article}
\usepackage{graphicx}
 \usepackage{amsmath}
 \usepackage{amsfonts}
 \usepackage{mathtext}

\begin{document}



\centerline{\Large \bf
Can bimodality exist without phase transition?}

\vspace*{1.1cm}

\centerline{\large \bf
V. V. Sagun, A. I. Ivanytskyi,  D. R. Oliinychenko,  K. A. Bugaev
}

\abstract{
\it Here we present an explicit counterexample to the widely spread beliefs 
about  an exclusive role of bimodality as the first order phase transition  signal. 
On the basis of an exactly solvable statistical model generalizing the  statistical multifragmentation model of nuclei   we demonstrate  that the bimodal nuclear fragment size distributions  can naturally  appear  in  infinite system without a phase transition. It appears at the supercritical temperatures due to the negative values of the surface tension coefficient. The developed statistical  model  corresponds to the compressible nuclear liquid  with the tricritical endpoint  located at one third of the normal nuclear density.
}

\section{Introduction}
Nowadays the  bimodality is considered as an unambiguous signal of the first order phase transition  (PT) in finite but large  systems. The authors  of such beliefs   \cite{THill:1,Bmodal:Chomaz03,Bmodal:Gulm07}  identify each  local maximum of the bimodal distribution with a pure phase. For instance, T. Hill justified  his assumption on bimodality appearance in finite systems  by stating that due to the fact that an interface between two pure phases `costs' some additional energy, the probability of their coexisting in a finite system is less than for each of pure phases \cite{THill:1}. At the same time  it is believed 
\cite{THill:1,Bmodal:Chomaz03,Bmodal:Gulm07} that  in the thermodynamic limit
a bimodality corresponds to a mixed phase only. 

Here we give an explicit counterexample based on the exact analytical solution of the constrained statistical multifragmentation model (CSMM) of nuclei  in the thermodynamic limit which leads to the bimodal fragment distributions inside of  the cross-over  region without the phase transition existence. In addition, we develop  a  realistic equation of state for the liquid phase which, in contrast to the original SMM formulation \cite{SMM:Bugaev00}, is a compressible one.  The suggested approach obeys the L. van Hove axioms of statistical mechanics. The second important element of the present model  is a  realistic  parameterization of the surface tension temperature dependence  which is  based on the exact analytical solution of the  partition function of surface deformations.

\section{CSMM with compressible nuclear liquid in thermodynamic limit}

The general solution of the CSMM  partition function formulated in the grand canonical variables of volume $V$, temperature $T$ and baryonic chemical potential $\mu$ is given by \cite{Bugaev:CSMM05}
\begin{equation}\label{EqI}
{\cal Z}(V,T,\mu)~ = \sum_{\{\lambda _n\}}
e^{\textstyle  \lambda _n\, V }
{\textstyle \left[1 - \frac{\partial {\cal F}(V,\lambda _n)}{\partial \lambda _n} \right]^{-1} } \,,
\end{equation}
where  the set of  $\lambda_n$ $(n=0,1,2, 3,..)$ are all the complex roots of  the equation 
\begin{equation}\label{EqII}
\lambda _n~ = ~{\cal F}(V,\lambda _n)\,,
\end{equation}
ordered as   $Re(\lambda_n) > Re(\lambda_{n+1})$ and $Im (\lambda_0) = 0$. The function ${\cal F}(V,\lambda)$ is defined as 
\begin{eqnarray}\label{EqIII}
&&\hspace*{-0.6cm}{\cal F}(V,\lambda)=\left(\frac{m T}{2 \pi}\right)^{\frac{3}{2}}z_1\exp \left\{\frac{\mu-\lambda T b}{T}\right\}+\hspace*{-0.1cm} \sum_{k=2}^{K(V)}\phi_k (T) \exp \left\{\frac{( p_l(T,\mu)- \lambda T)b k }{T} \right\}\,.~~~~~
\end{eqnarray}
Here $m \simeq 940$ MeV is a nucleon mass, $z_1 = 4$ is an internal partition (the degeneracy factor) of nucleons, $b = 1/ \rho_0 $ is the eigen volume of one nucleon in a vacuum ($\rho_0\simeq 0.17$ fm$^3$ is the normal nuclear density at $T=0$ and zero pressure). The reduced distribution function of the $k$-nucleon fragment in (\ref{EqIII}) is defined as 
\begin{equation}\label{EqIV}
 \phi_{k>1}(T)\equiv\left(\frac{m T }{2 \pi}\right)^{\frac{3}{2} } k^{-\tau}\,\exp \left[ -\frac{\sigma (T)~ k^{\varsigma}}{T}\right]\,,
\end{equation} 
where $\tau \simeq 1.825$ is the Fisher topological exponent and $\sigma (T)$ is the $T$-dependent surface tension coefficient.  Usually, the constant, parameterizing the dimension of surface in terms of the volume is  $\varsigma = \frac{2}{3}$.

In (\ref{EqIII}) the exponents $\exp( - \lambda b k)$ ($k=1,2,3,...$) appear due to the hard-core  repulsion between the nuclear fragments \cite{Bugaev:CSMM05}, while $p_l(T,\mu)$ is the pressure of the liquid phase. 

We consider the thermodynamic limit only, i.e. for $V \rightarrow \infty$  it follows $K(V) \rightarrow \infty$. Then the treatment of the model is essentially simplified, since Eq. (\ref{EqII}) can have only two kinds of solutions \cite{Bugaev:CSMM05}, either the gaseous pole $p_g (T, \mu) = T \lambda_0 (T, \mu)$ for  ${\cal F}(V,\lambda_0 - 0) < \infty$ or the liquid essential singularity 
$p_l (T, \mu) = T \lambda_0 (T, \mu)$ for  ${\cal F}(V,\lambda_0 - 0) \rightarrow  \infty$. The mathematical reason why  only the rightmost solution   $\lambda_0 (T, \mu) = \max \{Re(\lambda_n)\} $  of   Eq. (\ref{EqII}) 
 defines the system pressure is evident from Eq. (\ref{EqI}): in the limit $V \rightarrow \infty$ all  the solutions  of (\ref{EqII}) other than the rightmost one are exponentially suppressed.  

\begin{figure}[h]
\begin{minipage}[h]{0.49\linewidth}
\center{\includegraphics[width=1.0\linewidth]{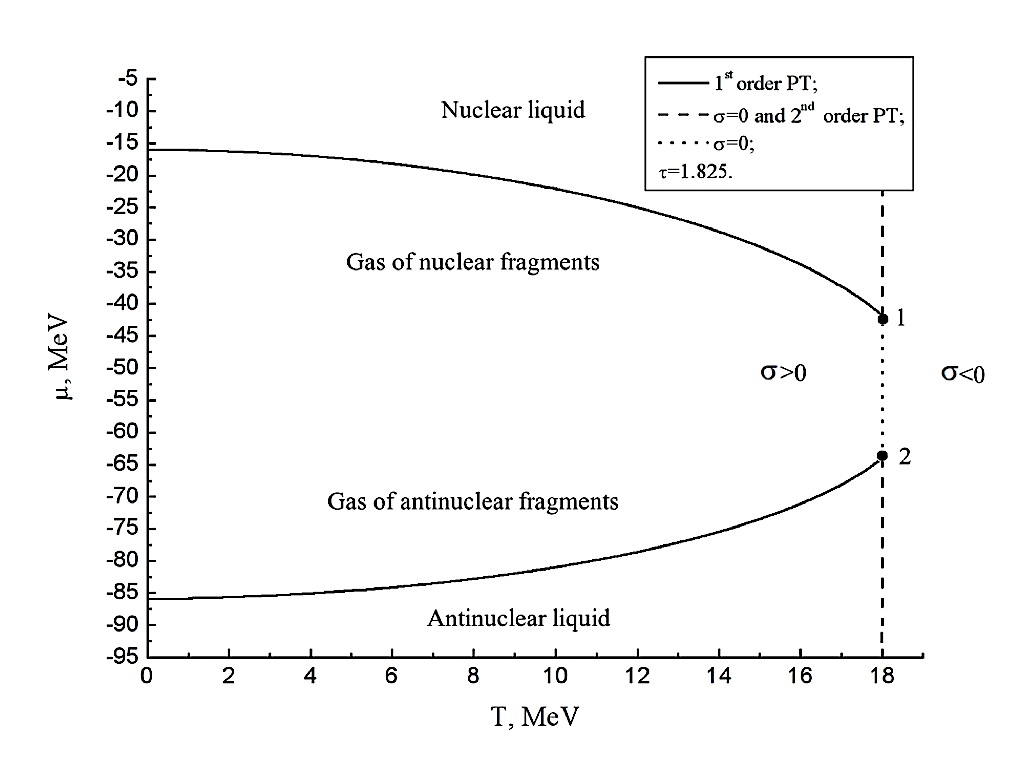} \\ a)}
\end{minipage}
\hfill
\begin{minipage}[h]{0.49\linewidth}
\center{\includegraphics[width=1.0\linewidth]{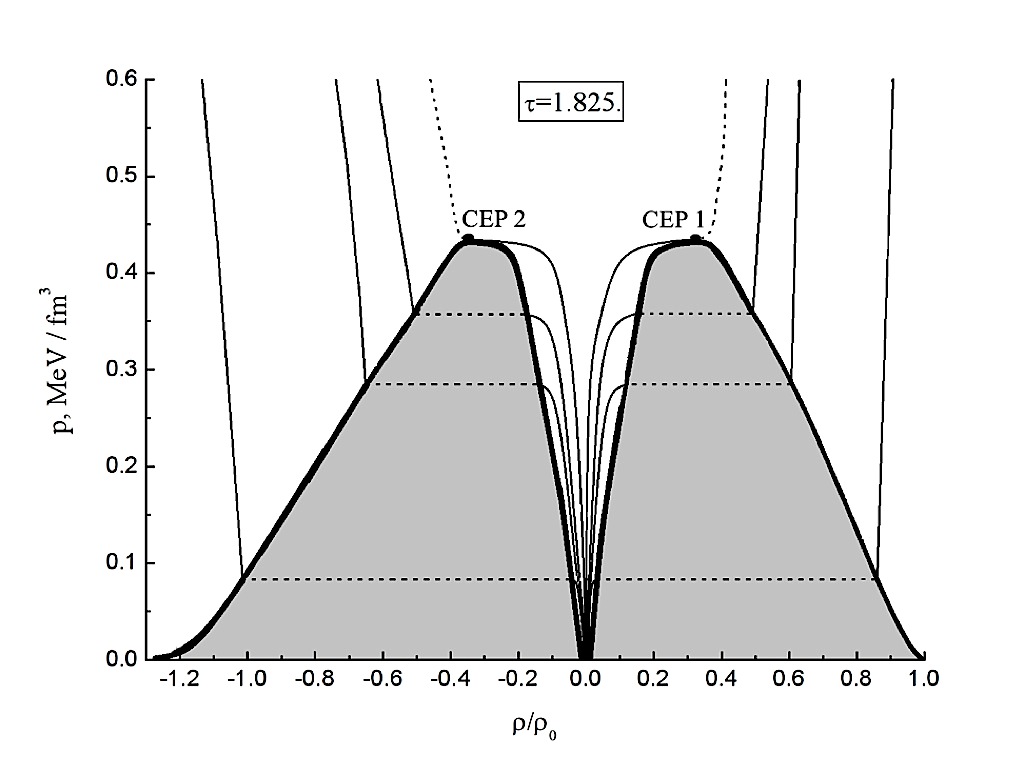} \\ b)}
\end{minipage}
\caption{Model phase diagrams are shown  in $T-\mu$ plane (a)) and $\rho-p$ plane (b)). In  the panel a)  a first order PT is shown by  the solid curves. The vertical dashed lines display the  second order PTs  and the black circles correspond to the tricritical endpoints marked  as 1 (nuclear matter) and 2 (antinuclear matter). A cross-over occurs along the  dotted vertical line of the vanishing surface tension coefficient.  The grey  areas in the  panel  b)  show the mixed phases of the first  order PTs. The isotherms are shown for $T=11, 16, 17, 18$ MeV  from bottom to top. Negative baryonic charge densities correspond to an   antimatter.  }
\label{ris:image1}
\end{figure}

In the thermodynamic limit the model has a PT, when there occurs a change of the  rightmost solution type, i.e. when the gaseous pole is changed by  the liquid essential singularity or vice versa. The PT line $\mu = \mu_c (T)$ is a solution of  the equation of  `colliding singularities' $p_g (T, \mu) = p_l (T, \mu) $, which is just the Gibbs criterion of  phase equilibrium. The properties of a PT are defined only by the liquid phase pressure  $p_l (T, \mu)$ and   by the temperature dependence of  surface tension $\sigma(T)$.

In order to avoid the incompressibility of the nuclear liquid we suggest to employ  the following  parameterization of its pressure 
\begin{eqnarray}
\label{EqV}
p_l=\frac{ W(T) +  \mu + a_2 ( \mu -\mu_0)^{2} + a_4 ( \mu -\mu_0)^{4}}{b} \,.
\end{eqnarray}
Where $ W(T) = W_0 + \frac{T^2}{W_0}$ denotes  the usual  temperature dependent  binding energy per nucleon with $W_0 =  16$ MeV,  while the constants  $\mu_0 = - W_0$, $a_2 \simeq 1.233 \cdot 10^{-2}$ MeV$^{-1}$ and $a_4 \simeq 4.099 \cdot 10^{-7}$ MeV$^{-3}$  are  fixed by  the requirement   to reproduce  the  normal nuclear matter properties, i.e. at vanishing temperature  $T=0$ and normal nuclear density $\rho = \rho_0$ the liquid pressure must be zero. 

In addition to the new parameterization of the free energy of the $k$-nucleon fragment (\ref{EqIII}) we  propose to use  the following  parameterization of the surface tension coefficient 
\begin{equation}\label{EqIX}
 \sigma (T) =  \sigma_0 \left| \frac{T_{cep} - T }{T_{cep}} \right|^\zeta  {\rm sign} ( T_{cep} - T) ~,
\end{equation}
with  $\zeta = const \ge 1$, $T_{cep} =18$ MeV and $\sigma_0 = 18$ MeV the SMM. In contrast to the Fisher droplet model \cite{Fisher:67} and the usual SMM, the CSMM surface tension (\ref{EqIX}) is negative above the critical temperature $T_{cep}$.

\begin{figure}[h]
\hfil\includegraphics[height=3.1in]{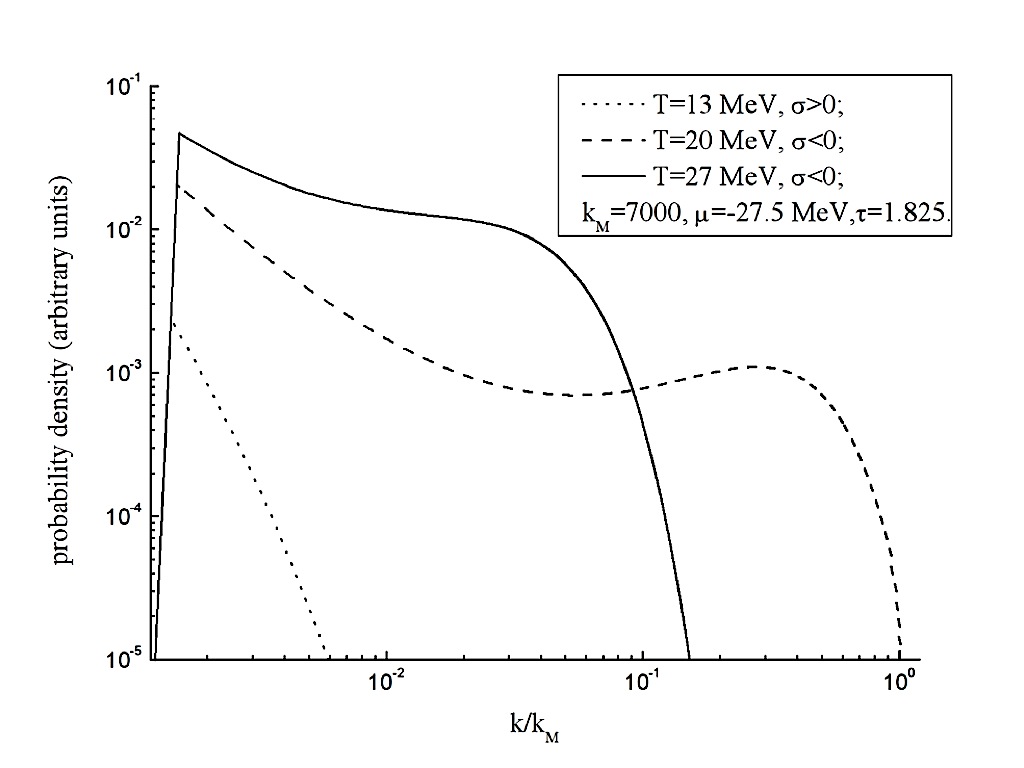}\hfil
\caption
{
  Fragment size distribution in the gaseous phase  is shown for a fixed baryonic chemical potential $\mu = -27.5$ MeV and three values of the  temperature $T$. The dotted curve is found exactly  at the boundary of gaseous and mixed phases.
}
\label{fig4}
\end{figure}

Now we would like to study the fragment size distribution in two regions  of the phase diagram in order to elucidate the role of the negative surface tension coefficient.  In order to demonstrate the pitfalls of the bimodal concept of Refs. \cite{Bmodal:Chomaz03, Bmodal:Gulm07,THill:1} we study the gaseous phase and the supercritical temperature region, where there is  no  PT  by construction.  As one can see from Fig. \ref{fig4} in the gaseous phase, even at the boundary with the mixed phase,  the size distribution is a monotonically   decreasing function of  the number of nucleons $k$ in a fragment. However, for the supercritical temperatures one finds the typical bimodal fragment size distribution for a variety of temperatures and chemical potentials as one can see from Fig.\ref{fig4}.

A sharp peak at low $k$ values reflects  a fast increase of the  probability density of dimers compared to the monomers (nucleons),  since the intermediate  fragment sizes do not have the binding free energy and the surface free energy and, hence, the monomers are significantly suppressed in this region of thermodynamic parameters. On the other hand, it is clear that the tail of fragment distributions in Fig. \ref{fig4} decreases due to the dominance of the bulk free energy and, hence, the whole structure at intermediate fragment sizes is due to  a competition between the surface free energy and two other contributions into the fragment free energy, i.e. the bulk one and the Fisher one.

\section{Conclusions}
In the present work we showed  that the bimodal distributions can naturally  appear in  infinite system without a PT. 
At  the supercritical temperatures
a bimodal distribution is generated  by the negative values of the surface tension coefficient. This result is in line with the previously discussed role of  the competition between the volume and the surface parts of  the system free energy.

Also we  suggested  the new parameterization of the CSMM liquid phase pressure which repairs the two main pitfalls of the original SMM and allows one to consider the compressible nuclear liquid which has the tricritical endpoint at the one third of the normal nuclear density. Surprisingly,   the suggested approach  to account for  the nuclear liquid compressibility automatically leads to   an appearance of  an  additional state that in many respects resembles the physical antinuclear matter.

\end{document}